\documentclass[12pt]{iopart}

\usepackage{graphicx}
\def\lesssim{\mathrel{\hbox{\rlap{\hbox{\lower4pt\hbox{$\sim$}}}\hbox{$<$}}}}
\def\gtrsim{\mathrel{\hbox{\rlap{\hbox{\lower4pt\hbox{$\sim$}}}\hbox{$>$}}}}

\begin{document}

\letter{A quantum-mechanical study of rotational transitions in H$_{2}$ induced by H}

\author{S A Wrathmall and D R Flower}
\address{Department of Physics, University of Durham, Durham, DH1 3LE, UK}
\eads{\mailto{s.a.wrathmall@dur.ac.uk}, \mailto{david.flower@dur.ac.uk}}

\begin{abstract}
Cross sections have been computed for rotational transitions of H$_2$, induced by collisions with H atoms, using a recent H -- H$_2$ potential calculated by Mielke et al. \cite{Mielke}. These results are compared with those obtained with earlier potentials. Significant discrepancies are found with results deriving from the potential of Boothroyd et al. \cite{BKMP} in the low collision energy regime. We compare also cross sections derived using different levels of approximation to the vibrational motion.
\end{abstract}

\bigskip

The H -- H$_2$ atom--molecule system is the simplest triatomic system; but, in spite of its structural simplicity, it has proved difficult to calculate the interaction potential to an accuracy which is sufficient to determine reliably the rotational excitation cross sections near threshold. In the regime of collision energies $E \lesssim 5000$~K, a potential barrier inhibits H--atom exchange scattering, and rotational excitation of the molecule occurs predominantly through non--exchange scattering; at higher energies, exchange--scattering occurs also. These processes contribute to the excitation of H$_2$ in the interstellar medium, notably in jets associated with star formation. Pure rotational transitions of H$_2$ have been observed from levels up to $v = 0, J = 27$ (Rosenthal et al. \cite{Rosenthal}) by means of the {\it Infrared Space Observatory}. 

A recent study of the H -- H$_2$ potential by Mielke et al. \cite{Mielke} appears to have improved on the accuracy of earlier work by Boothroyd et al. \cite{BKMP} and Partridge et al. \cite{Partridge}. Whilst all these calculations agree with regard to the general features of the potential, which becomes repulsive at short range and displays only a shallow Van der Waals minimum of approximately 2~meV (20~K), there remain significant differences in the predicted anisotropy of the potential; and it is the anisotropy which determines the magnitude of the rotationally inelastic cross sections. In this Letter, we present the results of calculations of cross sections for rotational transitions within the $v = 0$ vibrational ground state of H$_2$, comparing the values obtained using different determinations of the H -- H$_2$ interaction potential.

\begin{figure}
\begin{centering}
\includegraphics{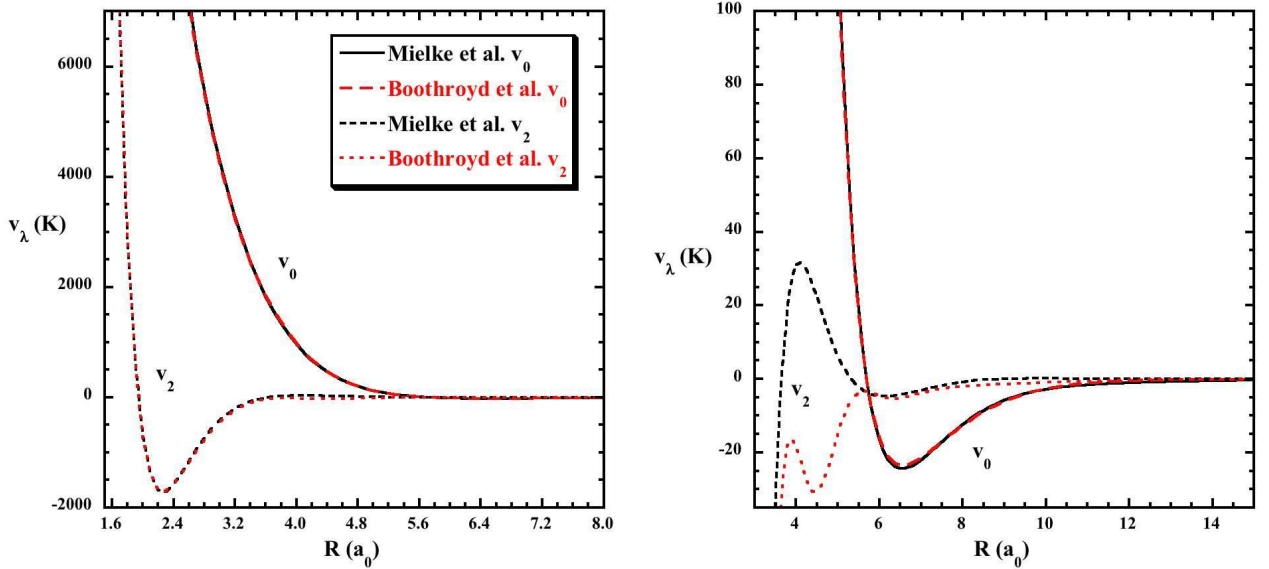}
\caption{The isotropic $v_0$ and first anisotropic $v_2$ potential expansion coefficients for the Boothroyd et al. and Mielke et al. potentials, computed in both cases for the equilibrium internuclear separation, $r = 1.401$~$a_0$, of the H$_2$ molecule in its ground vibrational state. The unit of the potential is K; $1 E_{\rm h} \equiv 315 780$~K.}
\label{potential}
\end{centering}
\end{figure}

When determining the rotational excitation cross sections, it is customary to expand the interaction potential $V(R,\theta ,r)$, in terms of Legendre polynomials, $P_{\lambda}(\cos\theta)$,
\begin{equation}
V(R,\theta ,r) = \sum^{\infty}_{\lambda=0} v_{\lambda}(R, r)P_{\lambda}(\cos\theta),
\label{potentialexp}
\end{equation}
where ${\bf R}$ is the vector connecting the centre of mass of the H$_2$ molecule to the H atom, ${\bf r}$ is the vector connecting the nuclei of the molecule, and $\cos \theta = {\bf \hat{R} \cdot \hat{r}}$. As H$_2$ is homonuclear, only those terms in the expansion with $\lambda $ even are non-zero. The results in Figure~\ref{potential} were obtained using a least--squares minimization procedure, which provides an excellent fit to the original potentials when terms through $\lambda = 10$ are included in the expansion.

The behaviour of the term $v_2 \left(R, r\right)$ in the classically allowed regions, where $E \gtrsim v_0 \left(R, r\right)$, essentially determines the magnitude of the $\Delta J = 2$ rotationally inelastic cross sections, in the range of collision energies considered here. Figure~\ref{potential} shows the isotropic term, $v_0$, and first non-zero anisotropic term, $v_2$, derived from the potential energies calculated by Mielke et al. \cite{Mielke} and Boothroyd et al. \cite{BKMP} for the equilibrium internuclear distance, $r = 1.401$~$a_0$, of the H$_2$ molecule in its ground vibrational state, $v = 0$. There is excellent agreement for the isotropic term, $v_0$, but, as the expanded plot to the right in Figure~\ref{potential} shows, there are important differences for the first non-zero anisotropic term, $v_2$. The calculations of Boothroyd et al. imply a $v_2$ term which remains negative beyond its minimum, at $R = 2.25$ $a_0$, whereas the calculations of Mielke et al. lead to a $v_2$ term which varies between positive and negative values for $3.7 \le R \le 12.9$ $a_0$. For atom--molecule separations, $R$, which correspond to low classically--allowed energies, the $v_2$ term is small in magnitude for both potentials. Therefore, we expect the rotationally inelastic cross sections to be small at low collision energies. This expectation is confirmed by the results reported below. It is evidently a difficult task to determine reliably the value of the anisotropic term when its magnitude is as small as shown in the right hand panel of Figure~\ref{potential}. However, we note that, in the range of $R$ which concerns us here, the $v_2$ coefficient which derives from the calculations of Mielke et al. is in distinctly better agreement with the 1993 results of Partridge et al. \cite{Partridge} than with the 1996 results of Boothroyd et al. \cite {BKMP}.

As the collision energy increases and the distance of closest approach decreases, the magnitude of the $v_2$ coefficient increases and we find better agreement between the results deriving from the Mielke at el. and Boothroyd et al. potentials. Thus, we expect the rotationally inelastic cross sections to increase in magnitude, and the two potentials to yield similar results. Once again, we confirm  this expectation below.

\begin{figure}
\begin{centering}
\includegraphics{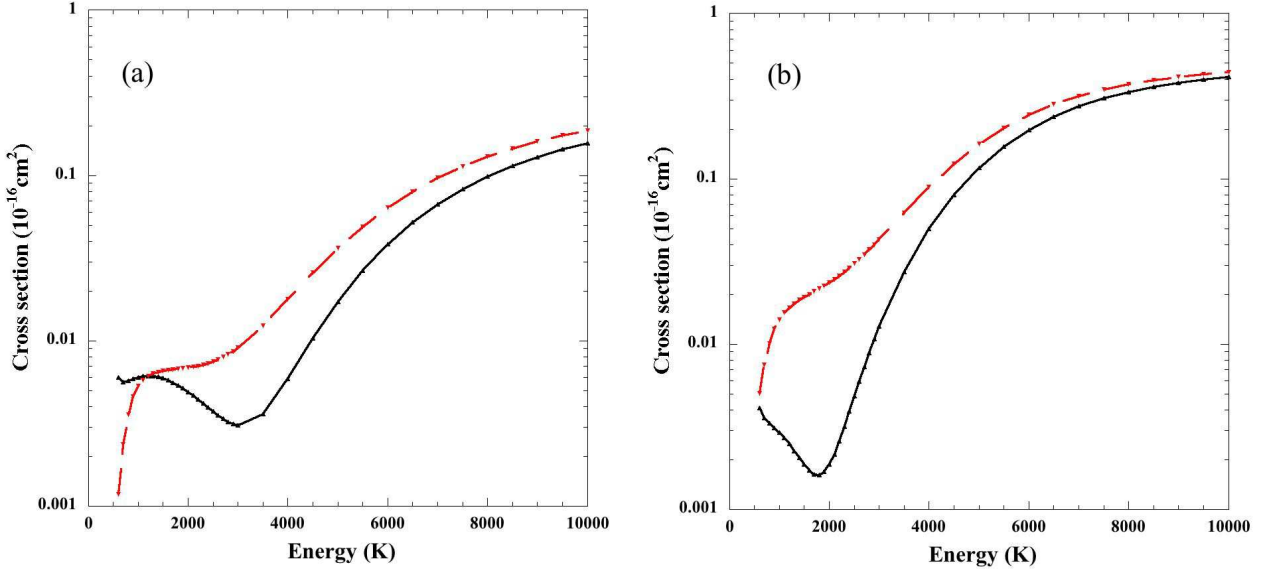}
\caption{Cross sections for the $J = 2 \rightarrow 0$ transition, in units of $10^{-16}$ cm$^2$. The centre--of--mass collision energy is expressed in K, relative to the $J = 0$ level. The full curve was obtained with the potential of Mielke et al., and the broken curve with the potential of Boothroyd et al.: (a) at the equilibrium internuclear distance, $r = 1.401$~$a_0$, in the vibrational ground state of the H$_2$ molecule; (b) using the expectation values of $v_{\lambda}\left(R, r\right)$ in the vibrational ground state (Equation~(\ref{vibrating})).}
\label{paracombined}
\end{centering}
\end{figure}

In addition to calculating the cross sections with the internuclear separation fixed at its equilibrium value, $r = 1.401$~$a_0$, in the vibrational ground state, we adopted a more exact approach, consisting of evaluating 

\begin{equation}
v_{\lambda}\left(R\right) = \int_{r_1}^{r_2}\psi^{*}\left(r\right)v_{\lambda}\left(R, r\right)\psi\left(r\right)dr\label{vibrating}
\end{equation}
where $\psi\left(r\right)/r$, the vibrational ground state eigenfunction at internuclear separation $r$, differs significantly from zero for $r_1 \le r \le r_2$. The vibrational eigenfunction was determined using the method of Marston and Balint-Kurti \cite{BalintExact} and the potential of Mielke et al. or Boothroyd et al., as appropriate.

Rotationally inelastic cross sections were computed using the MOLCOL code \cite{MOLCOL} and the exact coupled channels (CC) method \cite{Launay}. In order to ensure convergence of the cross sections, for the range of collision energies considered in this Letter, we included the first 18 rotational states of the H$_2$ molecule, $0 \leq J \leq17$, in the basis set. The $J = 17$ level lies at an energy of 21412~K above $J = 0$. Because hydrogen--atom exchange is assumed not to occur, the ortho and para forms of H$_2$ remain distinct, and separate calculations were performed including either the levels with $J$ odd or the levels with $J$ even. The first six terms $\left(\lambda = 0, 2, 4, 6, 8, 10\right)$ in the expansion (\ref{potential}) of the potential were used in the calculations.

In Figure~\ref{paracombined}a, we compare the cross section for de-excitation, $J = 2 \rightarrow 0$, as calculated with the potentials of Mielke et al. \cite{Mielke} and Boothroyd et al. \cite{BKMP}, for the case of a fixed internuclear separation, $r = 1.401$~$a_0$. As anticipated, the cross-section tends to increase with the collision energy,  and results obtained with the two potentials come into agreement. However as the collision energy decreases, the results obtained with the potential of Mielke et al. fall increasingly below those derived from the other potential, before rising again towards the $J = 2$ threshold. These differences relate to to the $v_2$ term in the potential expansion (see Figure~\ref{potential}). We have verified that the cross sections shown in Figure~\ref{paracombined}a are reproducible by the independent scattering code MOLSCAT \cite{MOLSCAT}, to within the plotting accuracy.

\begin{figure}
\begin{centering}
\includegraphics{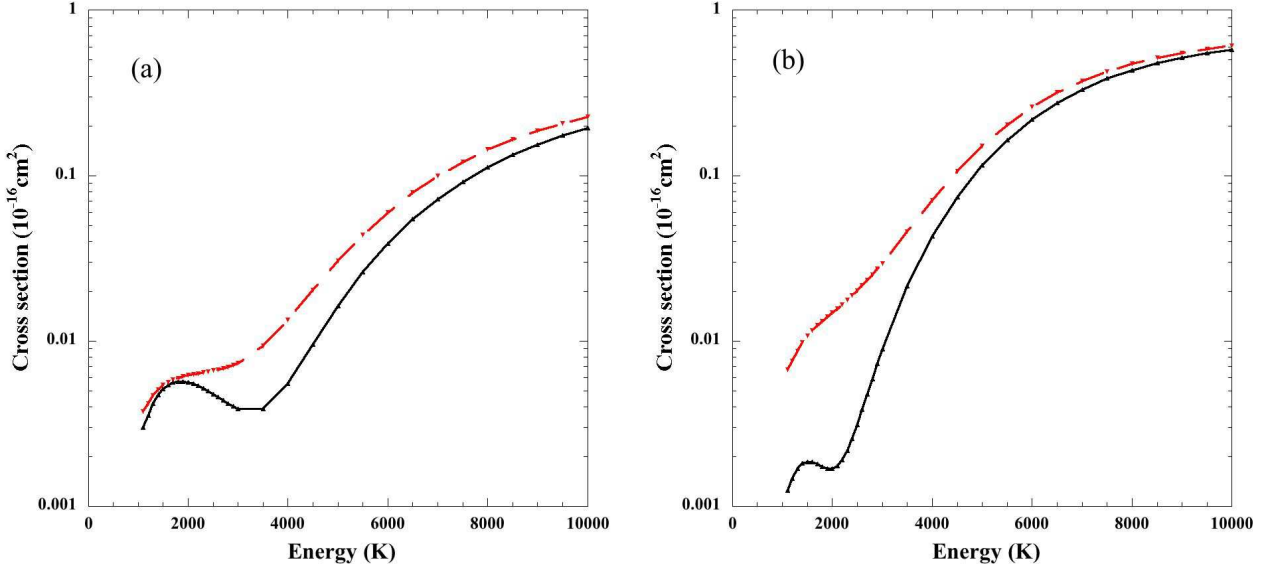}
\caption{As for Figure~\ref{paracombined}, but for the $J = 3 \rightarrow 1$ transition. The centre--of--mass collision energy is expressed in K, relative to the $J = 0$ level. Cross sections obtained using $v_{\lambda}(R, r = 1.401\ a_0)$ are shown in panel (a), and those derived using Equation~(\ref{vibrating}) in panel (b).}
\label{orthocombined}
\end{centering}
\end{figure}

Figure~\ref{orthocombined}a shows the corresponding results for the $J = 3 \rightarrow 1$, transition of ortho--H$_2$. The behaviour of this cross section is similar to that seen in Figure~\ref{paracombined}a, for the equivalent transition in para--H$_2$, except close to threshold ($E = 1015$~K), where both potentials yield a cross section which decreases with $E$. Once again, these results have been verified using MOLSCAT.

In Figures~\ref{paracombined}b and \ref{orthocombined}b, the rotational de-excitation cross sections for para-- and ortho--H$_2$ are compared for the case in which Equation~(\ref{vibrating}) was used to evaluate the potential expansion coefficients. Results obtained in this way should be more accurate than those derived assuming the internuclear distance, $r$, to be fixed at its equilibrium value. We see features in the cross sections which are similar to those apparent in Figures~\ref{paracombined}a and \ref{orthocombined}a. The discrepancies between the results obtained using different potentials are even greater than previously for collision energies $E \lesssim 3000$~K. Additional calculations were performed for models in which simple harmonic oscillator and Morse oscillator eigenfunctions were used in Equation~(\ref{vibrating}); they yielded results which are intermediate between those in Figures~\ref{paracombined} and \ref{orthocombined} and show the same features.

As mentioned above, the behaviour with collision energy and the magnitudes of the cross sections relate to the $v_2$ term in the expansion of the interaction potential. For scattering at low energies, the region $R \gtrsim 4$ $a_0$ is relevant, where the $v_2$ coefficient calculated by Mielke et al. \cite{Mielke} is in distinctly better agreement with that of Partridge et al. \cite{Partridge} than that of Boothroyd et al. \cite{BKMP}. Consequently, it was not surprising to find that the cross sections which derive from potentials \cite{Mielke} and \cite{Partridge} are also in better agreement. The rms and largest errors in the interaction energies computed by Boothroyd et al. (85.59~K and 1958.57~K, respectively) and Mielke et al. (3.25~K and 87.61~K, respectively) are to be compared with $|v_2| << 100$~K for $R \gtrsim 4$ $a_0$. Thus, the rms error quoted for surface of Mielke et al. \cite{Mielke} approaches that required to calculate reliably the cross sections for at least the $\Delta J = 2$ rotationally inelastic transitions. In the absence of any independent test of the accuracy of the calculated interaction potentials, it is clear that the cross sections computed with the potential of Mielke et al. are to be preferred.

We note that the low--energy cross sections which derive from the potential surface of Mielke et al. \cite{Mielke} are even smaller than those obtained using the surface of Boothroyd et al., with values which fall to the order of $10^{-19}$ cm$^2$ for $E \approx 2000$~K; the corresponding thermal rate coefficient is of the order $10^{-13}$ cm$^3$ s$^{-1}$. This value is 2 orders of magnitude smaller than the rate coefficients for the $J = 2 \rightarrow 0$ and $J = 3 \rightarrow 1$ transitions in H$_2$, induced by collisions with He atoms at $T = 2000$~K \cite{He}. In the interstellar medium, the elemental number density ratio of helium to hydrogen is $n_{\rm He}/n_{\rm H} = 0.1$. Thus, even if the hydrogen is mainly in atomic rather than molecular form, collisions with helium atoms will dominate the rotational excitation of the residual molecular hydrogen, owing to the much larger values of the He--H$_2$ rate coefficients.

\ack{SAW is supported by a research studentship from the Particle Physics and Astronomy Research Council.}


\section*{References}

\begin{thebibliography}{99}

\bibitem{Mielke}  Mielke S L Garrett B C Peterson K A 2002 \emph{J. Chem. Phys.} \textbf{116} 4142 
\bibitem{Rosenthal}  Rosenthal D Bertoldi F Drapatz S 2000 \emph{Astron. Astophys.} \textbf{356} 705 
\bibitem{BKMP}  Boothroyd A I Keogh W K Martin P G Peterson M R 1996 \emph{J. Chem. Phys.} \textbf{104} 7139
\bibitem{Partridge}  Partridge H Bauschlicher C W Stallcop J R Levin E 1993 \emph{J. Chem. Phys.} \textbf{99} 5951
\bibitem{BalintExact} Marston C C Balint-Kurti G G 1989 \emph{J. Chem. Phys.} \textbf{91} 3571
\bibitem{MOLCOL} Flower D R Bourhis G Launay J-M 2000 \emph{Computer Physics Communications} \textbf{131} 187
\bibitem{Launay} Launay J -M 1977 \emph{J. Phys. B: At. Mol. Opt. Phys.} \textbf{10} 3665
\bibitem{MOLSCAT} Green J M Hutson S 1994 \emph{Collaborative Computational Project 6} Daresbury Laboratory : UK Science and Engineering Research Council                                                                                                                                                                                                                                                     
\bibitem{He} Flower D R Roueff E Zeippen C J 1998 \emph{J. Phys. B: At. Mol. Opt. Phys.} \textbf{31} 1105
\end{thebibliography}
\end{document}